\documentstyle[multicol,prb,aps,ifthen,epsfig]{revtex}

\newcommand{\wide}[2]{                                                        %
\end{multicols}                                                               %
\widetext                                                                     %
\noindent                                                                     %
\ifthenelse{\equal{#1}{t}}                                                    %
{}                                                                            %
{                                                                             %
\raisebox{0.1in}[0in][0.02in]{$\rule{3.575in}{0.002in}                        %
\rule{0.002in}{0.08in}$}                                                      %
}                                                                             %
#2                                                                            %
\ifthenelse{\equal{#1}{b}}                                                    %
{}                                                                            %
{                                                                             %
{\raisebox{-0.1in}[0in][0.02in]                                               %
{\hspace{3.575in}$\rule{0.002in}{0.08in}                                      %
\rule[0.08in]{3.575in}{0.002in}$}                                             %
}                                                                             %
}                                                                             %
\begin{multicols}{2}                                                          %
\noindent                                                                     %
}                                                                             %

\def  \tG     {\tilde{G}}
\def  \uG     {\underline{G}}

\def  \tH     {\tilde{H}}
\def  \tJ     {\tilde{J}}
\def  \tsig   {\tilde{\sigma}}
\def  \bsig    {\mbox{\boldmath$\sigma$}}
\def  \bssig    {\mbox{{\scriptsize \boldmath$\sigma$}}}
\def  \balph   {\mbox{\boldmath$\alpha$}}
\def  \bnab    {\mbox{\boldmath$\nabla$}}
\def  \bdel    {\mbox{\boldmath$\delta$}}
\def  \epsi     {\varepsilon}
\def  \lims     {\lim_{s\rightarrow0^+}}
\def  \bra      {\langle k,s|}
\def  \ket      {|k,s\rangle}
\def  \brakp    {\langle k',s|}
\def  \ketkp    {|k',s\rangle}
\def  \brakpp   {\langle k'',s|}
\def  \ketkpp   {|k'',s\rangle}
\def  \brasp    {\langle k,s'|}
\def  \ketsp    {|k,s'\rangle}
\def  \brakbs   {\langle\underline{k},-s|}
\def  \ketkbs   {|\underline{k},-s\rangle}
\def  \eff    {\mathit{eff}}
\def  \rxc    {\mathit{rxc}}
\def  \op     {{\bf p}}
\def  \no     {\nonumber}

\begin{document}

\title{Theory of the anomalous Hall effect from the Kubo formula and the Dirac equation}
\author{A. Cr\'epieux* and P. Bruno}
\address{Max-Planck-Institut f\"{u}r Mikrostrukturphysik, Weinberg
2, 06120 Halle, Germany}
\date{\today}
\maketitle

\begin{abstract}
A model to treat the anomalous Hall effect is developed. Based on
the Kubo formalism and on the Dirac equation, this model allows
the simultaneous calculation of the skew-scattering and side-jump
contributions to the anomalous Hall conductivity. The continuity
and the consistency with the weak-relativistic limit described by
the Pauli Hamiltonian is shown. For both approaches, Dirac and
Pauli, the Feynman diagrams, which lead to the skew-scattering
and the side-jump contributions, are underlined. In order to
illustrate this method, we apply it to a particular case: a
ferromagnetic bulk compound in the limit of weak-scattering and
free-electrons approximation. Explicit expressions for the
anomalous Hall conductivity for both skew-scattering and side-jump mechanisms are obtained.
Within this model, the recently predicted ``spin Hall effect'' appears naturally.  \\


\end{abstract}

\begin{multicols}{2}
\section{Introduction}
The Hall resistivity of magnetic materials, in addition to the
normal part proportional to the magnetic field, contains a
supplementary part proportional to the magnetization, called the
anomalous Hall resistivity
\begin{eqnarray}
  \rho_H=R_0H+R_SM,
\end{eqnarray}
where $R_0$ and $R_S$ are the normal and anomalous Hall
coefficients respectively, $H$ the magnetic field and $M$ the
magnetization. While the normal Hall effect results from the
Lorenz force, the anomalous Hall effect is due to the spin-orbit
coupling in the presence of spin polarization. Experimentally, the
normal and anomalous parts can be extracted by measuring the Hall
resistivity as a function of the magnetic field. At high magnetic
field, when the magnetic saturation is reached, we get a linear
variation of the Hall resistivity with a slope related to $R_0$
and an extrapolated value at zero magnetic field related to
$R_S$. The normal and anomalous Hall coefficients have been
determined for a large number of bulk alloys. These
studies\cite{Beitel58,Ashworth69,Majumdar77,Handley78,Sinha79}
reveal that the sign of $R_S$ can change according to the alloy
composition and that $|R_SM|$ is generally larger than $|R_0H|$
for typical values of the magnetic field.

For different reasons, renewed attention to the anomalous Hall
effect is observed quite recently. It is not only due to the
increasing interest in spin-dependent transport phenomena but
also because of some particular and interesting behaviors of the
anomalous Hall resistivity obtained experimentally in granular
alloys\cite{Denardin00}, in magnetic films\cite{Caulet99} and
multilayers\cite{SatoCanedy}. In addition, the anomalous Hall
effect is increasingly used as a measurement tool to detect for
example magnetization\cite{Ohno00}, dynamics of magnetic
domains\cite{Wunderlich} or perpendicular
anisotropy\cite{HaanNakagawa}. Besides, a new effect closely
related to the anomalous Hall effect, the ``spin Hall effect'',
has been recently predicted\cite{HirschZhang}.

In the sixties, a number of theoretical
works\cite{Karplus54,Smit,Luttinger58,Berger} attempted to
elucidate the physical mechanisms responsible for the anomalous
Hall effect and to calculate an explicit expression for the
anomalous Hall resistivity. A series of
controversies\cite{SmitBerger73,Lyo74,SmitBerger78} arose from
those pioneering works which were solved through detailed
calculations\cite{Nozieres73} and comparisons\cite{Lyo72}. It is
now accepted\cite{Chien80a} that two mechanisms are responsible
for the anomalous Hall effect: the skew-scattering proposed by
Smit\cite{Smit} and the side-jump proposed by
Berger\cite{Berger}.

\begin{figure} \centering
\epsfig{file=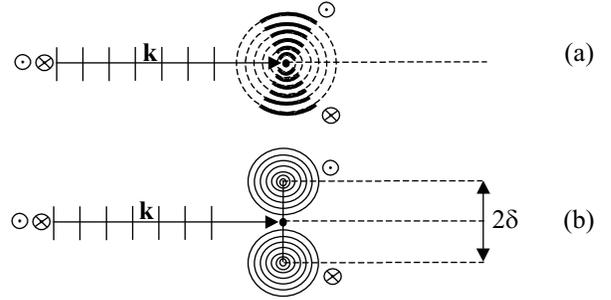}\vspace{0.25cm} \caption{Schematic
picture of the skew-scattering (a) and side-jump (b) mechanisms
from a quantum point of view ($\odot$ corresponds to spin up and
$\otimes$ to spin down). The bold curves represent the
anisotropic enhancement of the amplitude of the wave-packet due
to spin-orbit coupling.}\label{figQuan}
\end{figure}

An illustrative picture of these mechanisms is given in
Fig.~\ref{figQuan}. Consider an incident plane-wave characterized
by a wave-vector ${\bf k}$ which is scattered by a central
potential due, for example, to impurity. In the presence of
spin-orbit coupling, the amplitude of the wave-packet becomes
anisotropic in the sense that it depends of the relative
directions of the scattered and incident waves and of the spin.
After a succession of scattering events, the average trajectory
of the electron is deflected by a spin-dependent angle, which is
typically of order $10^{-2}$~rad. This first mechanism, depicted
by the diagram (a) in Fig.~\ref{figQuan}, corresponds to the
skew-scattering. The second mechanism corresponds to a lateral
displacement, $\delta\approx10^{-11}$~m, of the center of the
wave-packet during the scattering, which is also spin-dependent.
This mechanism, depicted by the diagram (b) in
Fig.~\ref{figQuan}, corresponds to the side-jump. In both cases,
due to the spin-orbit coupling, the effect is asymmetrical in
respect to the spin state. The spin up and spin down currents are
then different. In magnetic materials, this leads to a non-zero
spin current and to a transverse component in the charge current,
which corresponds to the anomalous Hall effect.

The skew-scattering and the side-jump mechanisms give different
contributions to the anomalous Hall resistivity. For bulk
material, it has been shown that, in certain limits, the
skew-scattering contribution is simply proportional to the
resistivity \cite{Smit,Luttinger58} while the side-jump
contribution is proportional to the square of the resistivity
\cite{Berger}. Then, we should have the simple expression
\begin{eqnarray}\label{anomres}
    \tilde{\rho}_H=\tilde{\rho}_{yx}=a\tilde{\rho}_{xx}+b\tilde{\rho}^2_{xx},
\end{eqnarray}
which implies that the relative importance of these two
contributions depends both on the temperature and on the impurity
concentration. However, we show in this paper that, even if the
relation (\ref{anomres}) remains correct, the skew-scattering
mechanism contributes also to the quadratic term in the case of
impurity scattering. Such behaviour was already been shown by
Kondorskii {\it et al.}\cite{Kondorskii}.

The traditional way to calculate the anomalous Hall resistivity
is to include the contribution of spin-orbit coupling in the
transition probability (it leads to the skew-scattering provided
one goes beyond the Born approximation) and in the velocity (it
leads to the so-called anomalous velocity which gives the
side-jump). While the skew-scattering can be obtained in a
classical approach it is claimed that the side-jump is a pure
quantum effect. We shall discuss this point in the Sec.~II of this
paper. Most of the calculations of the anomalous Hall resistivity
are based on the Bolztmann equation and used severe
approximations, in particular concerning the side-jump
contribution. Some calculations\cite{Kondorskii} are based on the
Kubo formalism, but surprisingly it is claimed that the side-jump
contribution vanishes, and only the skew-scattering contribution
is calculated.

Although the anomalous Hall effect is an old phenomena which has
motivated a lot of experimental and theoretical studies, a
unified model, able to calculate the skew-scattering and side-jump
contributions on the same footing, was still missing. In this
paper, we propose such a model. It is based on the Kubo formalism
and has the peculiarity to be built from the Dirac equation. The
justification for such an approach is given in Sec.~III where we
discuss in detail two different approaches for solving the
anomalous Hall effect i.e., based on Dirac and Pauli equations,
and study the consistency in the weak-relativistic limit of the
expressions of the conductivity tensors obtained in these two
approaches. In Sec.~IV, we calculate the anomalous Hall
conductivity of a disordered ferromagnetic bulk compound. The
results are discussed in Sec.~V.

\section{Comments on the physical nature of the side-jump mechanism}
It is often believed that the side-jump is a pure quantum effect
and has no classical equivalent\cite{Chien80a}. The usual
description of the side-jump is then based on a quantum picture
(see Fig.~\ref{figQuan}(b)) of a plane-wave transformed by
scattering in the presence of spin-orbit coupling into a spherical
wave whose center is shifted in a lateral direction
(perpendicular to the momentum and to the spin). The sign of the
displacement is opposite for spin up ($s=1$) and spin down
($s=-1$). A simple calculation in terms of phase-shift allows to
determine this displacement. We start from the Pauli Hamiltonian
\begin{eqnarray}\label{PauliH0}
    H&=&\frac{p^2}{2m}-\mu_B(\bsig\cdot{\bf B}_{\eff})+W=H_0+W,
\end{eqnarray}
where $\bsig$ is the Pauli matrix, ${\bf B}_{\eff}$ the effective
magnetic field due to exchange interactions and $W$ the total
potential including the spin-orbit coupling
\begin{eqnarray}\label{PauliW}
    W=V+\frac{\hbar}{4m^2c^2}(\bsig\times\bnab V)\cdot\op.
\end{eqnarray}
The state of the system $|\Psi_{ks}\rangle$ after scattering is
given in the Born approximation by the Lippmann-Schwinger equation
$|\Psi_{ks}\rangle = |k,s\rangle + \sum_{k's'} |k',s'\rangle
G_0({\bf k}',s',\epsi_k^s) \langle k',s'|W|k,s\rangle$, where
$\epsi_k^s$ and $G_0$ are respectively the eigenvalues and the
Green's function associated with $H_0$. The matrix elements of the
potential are
\begin{eqnarray}\label{mateleV}
  \langle k',s'|W|k,s\rangle=\tilde{V}_{\bf kk'}\left(\delta_{ss'}
  +\frac{i\hbar^2}{4m^2c^2}(\bsig_{s's}\times{\bf k}')\cdot{\bf
  k}\right),
\end{eqnarray}
where $\tilde{V}_{\bf kk'}$ is the Fourier transform of $V$. As
the spin-orbit term is imaginary, it will influence the phase of
the spherical wave. Thus, for small spin-orbit coupling, the wave
function $\Psi_{ks}({\bf r})=\langle r|\Psi_{ks}\rangle$ which
describes the wave after scattering can be expressed as
\begin{eqnarray}
 \Psi_{ks}({\bf r})\propto e^{i{\bf r}\cdot{\bf k}}+\sum_{k's'}\delta_{ss'}G_0({\bf
  k}'s',\epsi_k^s)\tilde{V}_{\bf kk'}e^{i{\bf r'_{s}}\cdot{\bf k}'},
\end{eqnarray}
where we have assumed that the effective magnetic field is along
the z-direction. The center of the wave packet after scattering
is given by
\begin{eqnarray}
 {\bf r}'_{s}\equiv{\bf r}+\frac{\hbar^2}{4m^2c^2}(\bsig_{ss}\times{\bf k}),
\end{eqnarray}
which is clearly spin dependent and means that the shift of the
center of the wave packet is different for spin up and spin down.
The lateral displacement, defined as $\bdel^s\equiv{\bf r}'_s-{\bf
r}$, is then equal to
\begin{eqnarray}\label{deltaquan}
 \bdel^s=\frac{\lambda^2}{4\hbar}(\bsig_{ss}\times\op),
\end{eqnarray}
where we have introduced the length $\lambda$ which corresponds
to the Compton wave length $\lambdabar_c=\hbar/mc$ in the case of
free electrons. In real materials, Berger\cite{Berger} has shown
that the spin-orbit coupling (i.e., $\lambda^2$) is renormalized
by band structure effects by a factor $\alpha\simeq10^4$. We then
obtain a lateral displacement $\bdel$ which is independent of
disorder and of order
$\lambda^2k_F/4\approx\alpha\lambdabar_c^2k_F/4\approx
10^{-11}$~m, in agreement with experimental results. Identical
expression for $\bdel$ was originally derived by Lyo {\it et
al.}\cite{Lyo72}.

\begin{figure} \centering
\epsfig{file=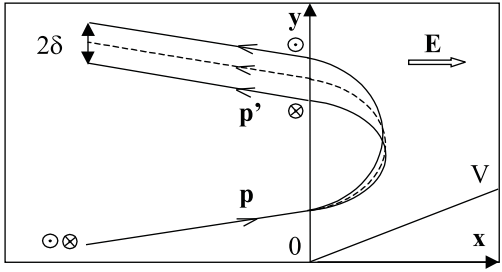}\vspace{0.25cm} \caption{Classical
picture of the side-jump mechanism. The dashed line correspond to
the non-relativistic trajectory of the particle and the solids
lines to the relativistic trajectories for spin up ($\odot$) and
spin down ($\otimes$).}\label{figClass}
\end{figure}

In a pure classical picture, such a lateral displacement can be
experienced by a particle with spin. As a simple example,
consider an electron with a charge $e$ ($e<0$) subject to a
uniform electric field ${\bf E}=E{\bf u}_x$ ($E>0$) in the region
$x>0$; there is no field in the region $x<0$. An incident electron
coming from the region $x<0$ is reflected by the field as
sketched in Fig.~\ref{figClass}.
The velocity is given by
\begin{eqnarray}
    {\bf v}=\frac{\partial H}{\partial\op}=\frac{\op}{m}-\frac{e\hbar}{4m^2c^2}(\bsig\times{\bf E}),
\end{eqnarray}
and therefore contains an anomalous contribution ${\bf
v}_a=-e\hbar(\bsig\times{\bf E})/4m^2c^2$ arising from the
spin-orbit interaction. In the field region ($x>0$), where the
trajectory is parabolic, the electron (we assume the spin to be
along the z-axis) has an anomalous velocity along the y-axis,
${v}_a^y=-e\hbar(\sigma_zE)/4m^2c^2$. The electron therefore
emerges with a shift along y, proportional to its spin
$\sigma_z$. For an arbitrary electric field, the shift due to the
anomalous velocity can be easily calculated
\begin{eqnarray}
    \bdel=\int_{-\infty}^{+\infty}{\bf v}_a\,dt=-\int_{-\infty}^{+\infty}\frac{e\hbar}{4m^2c^2}(\bsig\times{\bf E})\,dt,
\end{eqnarray}
with $e{\bf E}dt=d\op$, so that
\begin{eqnarray}\label{deltaclas}
    \bdel=-\frac{\hbar\bsig}{4m^2c^2}\times\int_{-\infty}^{+\infty}d\op=\frac{\lambda^2\bsig}{4\hbar}\times(\op-\op'),
\end{eqnarray}
In the above derivation, we have assumed that the spin is
perpendicular to the scattering plane. The lateral displacement
that we obtain is consistent with the one obtained in the quantum
picture. Indeed, the parallel can be simply done by replacing in
this classical calculation the momentum by a momentum operator
and by making the angular average over the final momentum $\op'$:
thus (\ref{deltaclas}) coincides with (\ref{deltaquan}).

\section{Comparison of the Dirac and Pauli approaches}
Generally, the calculations of the anomalous Hall conductivity are
based on the Pauli Hamiltonian. However, in our modelization,
i.e., within the framework of the Kubo formalism, it appears to be
simpler to adopt a relativistic approach based on the Dirac
equation. To justify that, let us first remember the derivation
of the skew-scattering and the site-jump contributions in the
Pauli approach. In presence of an exchange coupling, the Pauli
Hamiltonian is $H=\tH+H_{rc}$ where $\tH$ is the non-relativistic
Hamiltonian
\begin{eqnarray}\label{NR}
  \tH=\frac{p^2}{2m}-\mu_B\left(\bsig\cdot{\bf B}_{\eff}\right)+V,
\end{eqnarray}
and $H_{rc}$ the first relativistic corrections to the Hamiltonian
(order $1/c^2$)
\begin{eqnarray}\label{RCH}
  H_{rc}&=&-\frac{p^4}{8m^3c^2}+\frac{\hbar}{4m^2c^2}(\bsig\times\bnab V)\cdot{\bf p}\no \\
  &&+\frac{\hbar^2}{8m^2c^2}\Delta V+H_{\rxc},
\end{eqnarray}
which contains the relativistic mass correction, the spin-orbit
coupling, the Darwin term and the relativistic correction to the
exchange coupling $H_{\rxc}$. Since the effect we are interested
in results from the spin-orbit coupling, we do not need to give
the explicit expression of $H_{\rxc}$ (calculations and comments
on this term are presented in Ref. \onlinecite{Crepieux01}). In
this work, we do not consider the contribution of the periodic
part of the spin-orbit coupling (i.e., due to the lattice) but
only the aperiodic part due to the presence of impurities. In the
Pauli approach, the velocity contains two parts. One resulting
from the non-relativistic Hamiltonian ${\bf \tilde{v}}=\op/m$ and
another one resulting from the relativistic corrections
\begin{eqnarray}\label{RCV}
  {\bf v}_{rc}=-\frac{p^2\op}{2m^3c^2}+\frac{\hbar}{4m^2c^2}(\bsig\times\bnab V)+{\bf v}_{\rxc}.
\end{eqnarray}
where ${\bf v}_{\rxc}$ is the velocity related to $H_{\rxc}$. In
this description, the spin-orbit contribution to the velocity
(second term in (\ref{RCV})), the so-called anomalous velocity,
appears in a natural and transparent way. When we insert this
contribution in the Kubo formula, we obtain the side-jump
contribution. It is also possible to isolate the spin-orbit
contribution in the Green's function $G$ associated with $H$ by
making the following expansion
\begin{eqnarray}\label{Pauligreen}
  G=\tG+\tG H_{rc}\tG+\tG H_{rc}\tG H_{rc}\tG + ...,
\end{eqnarray}
where $\tG$ is the non-relativistic Green's function associated
with the non-relativistic Hamiltonian $\tH$. When we insert this
expression in the Kubo formula and proceed beyond the Born
approximation, we obtain the skew-scattering contribution.
Therefore, in the Pauli approach we get separately the
skew-scattering and the side-jump contributions when the Green's
functions and the velocities are respectively corrected by the
spin-orbit coupling.

One important problem in the Pauli approach is to treat disorder.
Actually, the spin-orbit coupling introduces two things:
off-diagonal disorder (in the tight-binding approximation) and
disorder in the velocity through the anomalous velocity. The
second consequence is critical because it is then difficult to
calculate precisely the vertex corrections and accordingly the
anomalous Hall resistivity. To avoid these problems, we have
chosen to base our model upon the Dirac equation instead of the
Pauli equation. In presence of an exchange coupling, it has the
form\cite{MacDonald79,Note1}
\begin{eqnarray}\label{Diracequ}
    H=c(\balph\cdot\op)+\beta mc^2+V-\mu_B\beta\left(\bsig\cdot{\bf B}_{\eff}\right),
\end{eqnarray}
where the first term is the kinetic energy, the second term the
mass energy, the third term is the potential and the last one the
exchange coupling. From (\ref{Diracequ}), we see that the
velocity is simply
\begin{eqnarray}\label{Diracvelo}
    {\bf v}=\frac{\partial H}{\partial \op}=c\balph
    =c\left(\begin{array}{clcr}
    0 & \bsig \\
    \bsig & 0
    \end{array}\right).
\end{eqnarray}
At this level, there appears an apparent contradiction between the
two approaches since, in the Dirac approach, contrary to the
Pauli approach, we do not have any spin-orbit contribution to the
velocity (anomalous velocity). It is therefore not clear {\it a
priori} whether the side-jump mechanism would emerge from the
Dirac approach. Actually, in the Dirac approach, the spin-orbit
coupling, although it does not appear explicitly, is properly
taken into account. Therefore, the conductivity should contain
simultaneously the skew-scattering and the side-jump
contributions as well as higher order contributions in $1/c^2$.
However, the expressions of the conductivity obtained in the
Dirac and Pauli approaches should coincide in the
weak-relativistic limit. To check this, we have calculated, in a
formal manner, the weak-relativistic limit up to order $1/c^2$ of
the conductivity obtained from the Dirac equation and compared it
with the conductivity obtained from the Pauli equation. The
determination of the conductivity tensor is performed in the Kubo
formalism. In certain limits, the conductivity can be expressed
as a product of operators, namely Green's functions and
velocities. However, the formulations proposed in the literature
are often confused or even
wrong\cite{Butler85,Banhart,Weinberger96} concerning the
off-diagonal elements of the conductivity tensor due to an abusive
generalization of the Kubo-Greenwood formula\cite{Greenwood58}.
In order to clarify the situation, we present in Appendix A the
derivation of the conductivity tensor from the original Kubo
formula\cite{Kubo} and summarize the different stages and
approximations which lead first to the Bastin
formula\cite{Bastin71} and finally to the Streda
formula\cite{Streda82}. We show that the latter is a sum of two
terms, $\tsig_{ij}^I$ and $\tsig_{ij}^{II}$ respectively given,
in the limits of independent electrons approximation, zero
temperature and zero frequency, by (\ref{streda4}) and
(\ref{streda5})
\begin{eqnarray}\label{stredatot}
  \tsig_{ij}&=&\tsig_{ij}^I+\tsig_{ij}^{II}, \no \\
  \tsig_{ij}^I&\equiv&\frac{e^2\hbar}{4\pi\Omega} {\mathrm Tr}\big\langle v_i(G^+-G^-)v_jG^--v_iG^+v_j(G^+-G^-)\big\rangle_c, \no \\
  \tsig_{ij}^{II}&\equiv&-\frac{e^2}{4i\pi\Omega}{\mathrm Tr}\big\langle(G^+-G^-)(r_iv_j-r_jv_i)\big\rangle_c,
\end{eqnarray}
where $i$ and $j$ are the direction indices, $\Omega$ the volume
of the sample, $\langle...\rangle_c$ denotes the configurational
average, $G^+$ and $G^-$ are the retarded and advanced Green's
functions at the Fermi level: $G^{\pm}=G(\epsi_F\pm
i0)=\left(\epsi_F\pm i0-H\right)^{-1}$. The procedure that we
follow is first to insert the Dirac velocity and Dirac Green's
function in (\ref{stredatot}), next to perform a weak-relativistic
expansion of $\tsig_{ij}$ and finally to compare it with the
expression obtained in the Pauli approach. The Dirac velocity is
given by (\ref{Diracvelo}) and for the Dirac Green's function, we
have used an exact expression derived from (\ref{Diracequ}) and
given in Ref.~\onlinecite{Crepieux01} by Eq.~(A3)
\wide{m}{
\begin{eqnarray}\label{Diracgreen}
  G^{\pm}=\left(\begin{array}{cc}
  \tG^{\pm}-\tG^{\pm}\frac{\bssig\cdot\op}{2mc}\big(\epsi_F-V-\mu_B\left(\bsig\cdot{\bf B}_{\eff}\right)\big)
  D^{\pm}\frac{\bssig\cdot\op}{2mc}\tG^{\pm}
  &\,\,\,\,\,\,\, \tG^{\pm}\frac{\bssig\cdot\op}{2mc} (Q^{\pm})^{-1}D^{\pm}Q^{\pm}\\
  D^{\pm}\frac{\bssig\cdot\op}{2mc}\tG^{\pm}
  &\,\,\,\,\,\,\, \frac{1}{2mc^2}D^{\pm}Q^{\pm}
  \end{array}\right),
\end{eqnarray}
}
where the operators $D^{\pm}$ and $Q^{\pm}$ are given by
\begin{eqnarray}\label{expD}
D^{\pm}&=&\left(1+Q^{\pm}\frac{\big(\epsi_F-V-\mu_B(\bsig\cdot{\bf B}_{\eff})\big)}{2mc^2}\right)^{-1}, \\
Q^{\pm}&=&1+\frac{(\bsig\cdot\op)\tG^{\pm}(\bsig\cdot\op)}{2m}.
\end{eqnarray}
The details of the calculations are presented in Appendix B. The
determination of the conductivity is done up to order $1/c^2$. It
is shown that the identification with the Pauli approach is
successful only when one considers the total conductivity
$\tsig_{ij}=\tsig^I_{ij}+\tsig^{II}_{ij}$. Indeed, when we
compare the expression of $\tsig^I_{ij}$ obtained in the Dirac
approach (see (\ref{sigmaI2}) for order $1/c^0$ and
(\ref{sigmaI3}) for order $1/c^2$) to the one obtained in Pauli
approach, we obtained different terms which are exactly canceled
by terms in $\tsig^{II}_{ij}$ (see (\ref{sigmaII4}) for order
$1/c^0$ and (\ref{sigmaII5}) for order $1/c^2$). The
non-relativistic limit of the total conductivity obtained in the
Dirac approach is
\begin{eqnarray}\label{TOT0}
  \tsig_{ij}^{(0)}&=&\frac{e^2\hbar}{4\pi\Omega}{\mathrm Tr}\left\langle \frac{p_i}{m}(\tG^+-\tG^-)\frac{p_j}{m}\tG^-\right. \no \\
  &&-\left.\frac{p_i}{m}\tG^+\frac{p_j}{m}(\tG^+-\tG^-)\right\rangle_c \no \\
  &&-\frac{e^2}{4i\pi\Omega}{\mathrm Tr}\left\langle(\tG^+-\tG^-)(r_i\frac{p_j}{m}-r_j\frac{p_i}{m})\right\rangle_c,
\end{eqnarray}
which corresponds exactly to the conductivity obtained from
(\ref{stredatot}) when one inserts the non-relativistic velocity
${\bf \tilde{v}}=\op/m$ and the non-relativistic Green's function
$\tG$. The last term in Eq.~(\ref{TOT0}) is zero in absence of
external magnetic field. The fact that a supplementary term in
$\tsig^{I(0)}_{ij}$ is present in the Dirac approch and not in the
Pauli approach has serious consequences when one neglects
$\tsig^{II(0)}_{ij}$ because it leads to an additional
contribution at order $1/c^0$ to the off-diagonal conductivity
which does not disappear in the non-relativistic limit and thus
would give unphysical results. At order $1/c^2$, the total
conductivity obtained in the Dirac approach is
\begin{eqnarray}\label{TOT2}
  &&\tsig_{ij}^{(2)}=\tsig_{ij}^{SS}+\tsig_{ij}^{SJ}+\tsig_{ij}^{or},
\end{eqnarray}
where $\tsig_{ij}^{SS}$ contains the terms which lead to the
skew-scattering
\begin{eqnarray}
  \tsig_{ij}^{SS}&=&
  \frac{e^2\hbar}{4\pi\Omega}{\mathrm Tr}\left\langle \frac{p_i}{m}(\tG^+H_{rc}\tG^+-\tG^-H_{rc}\tG^-)\frac{p_j}{m}\tG^-\right. \no \\
  & &+\frac{p_i}{m}(\tG^+-\tG^-)\frac{p_j}{m}\tG^-H_{rc}\tG^- \no \\
  & &-\frac{p_i}{m}\tG^+H_{rc}\tG^+\frac{p_j}{m}(\tG^+-\tG^-) \no \\
  &
  &-\left.\frac{p_i}{m}\tG^+\frac{p_j}{m}(\tG^+H_{rc}\tG^+-\tG^-H_{rc}\tG^-)\right\rangle_c,
\end{eqnarray}
$\tsig_{ij}^{SJ}$ contains the terms which lead to the side-jump
\begin{eqnarray}
  \tsig_{ij}^{SJ}&=&\frac{e^2\hbar}{4\pi\Omega}{\mathrm Tr}\left\langle({\bf v}_{rc})_i(\tG^+-\tG^-)\frac{p_j}{m}\tG^-\right. \no \\
  &-&({\bf v}_{rc})_i\tG^+\frac{p_j}{m}(\tG^+-\tG^-)+\frac{p_i}{m}(\tG^+-\tG^-)({\bf v}_{rc})_j\tG^- \no \\
  &-&\left.\frac{p_i}{m}\tG^+({\bf v}_{rc})_j(\tG^+-\tG^-)\right\rangle,
\end{eqnarray}
and $\tsig_{ij}^{or}$ is equal to
\begin{eqnarray}
  &&\tsig_{ij}^{or}=-\frac{e^2}{4i\pi\Omega}{\mathrm Tr}\left\langle
  \left(\tG^+-\tG^-\right)\left(r_i({\bf v}_{rc})_j-r_j({\bf v}_{rc})_i\right)\right. \no \\
  &&+\left.\left(\tG^+H_{rc}\tG^+-\tG^-H_{rc}\tG^-\right)
  \left(r_i\frac{p_j}{m}-r_j\frac{p_i}{m}\right)\right\rangle_c.
\end{eqnarray}
In addition to the skew-scattering and the side-jump
contributions to the anomalous Hall effect, we identify a new
contribution, $\tsig_{ij}^{or}$, which is related to the orbital
momentum ${\bf L}={\bf r}\times\op$. The expression (\ref{TOT2})
of the conductivity corresponds exactly to the one which is
obtained from (\ref{stredatot}) when one inserts the first order
corrections to the velocity ${\bf v}_{rc}$ and to the Green
function $\tG H_{rc}\tG$ where ${\bf v}_{rc}$ and $H_{rc}$ are
given respectively by Eqs. (\ref{RCH}) and (\ref{RCV}). We have
then proved in the weak-relativistic limit (up to order $1/c^2$)
the coincidence of the conductivity in the two approaches.

In summary, from the Pauli Hamiltonian, we get the
skew-scattering and the side-jump contributions separately while,
from the Dirac Hamiltonian, we get the both contributions and
also higher order in $1/c^2$ contributions simultaneously.
Therefore, in a full relativistic Dirac description, it will be
difficult to assess the importance of each contributions. However,
this approach has a great advantage over the Pauli approach: it
allows a simpler treatment of the disorder because, in contrast to
the Pauli approach where both the velocities and the Green's
functions contain disorder, the disorder is only present in the
Green's functions. It is thus possible to take one of the velocity
operator outside of the configurational average and to calculate
precisely the vertex corrections to the conductivity. For this
reason, the Dirac approach should be more efficient to calculate
the anomalous Hall resistivity.

In the next section, we present a direct application of our
model. In order to perform the analytical calculations, we
restrict ourselves to the weak-relativistic limit and to
approximate calculations of the vertex corrections; then the
results that we obtain can still be compared to the ones obtained
from the Pauli approach.

\section{Anomalous Hall conductivity of a ferromagnetic compound}
In this section, we present the calculation of the anomalous Hall
conductivity of a ferromagnetic bulk compound submitted to a
potential. This calculation is done in both Dirac and Pauli
approaches in order to show the similarities and the differences
between these two approaches. We consider a system with a cubic
symmetry and a magnetization along the z-axis. Thus, the
conductivity tensor has the form
\begin{eqnarray}\label{PauliH}
  \tsig=\left(\begin{array}{lll}
  \tsig_{xx} & \tsig_{xy} & 0 \\
  -\tsig_{xy} & \tsig_{xx} & 0 \\
  0 & 0 & \tsig_{zz}
  \end{array}\right).
\end{eqnarray}
We are only interested in the relativistic corrections to the
off-diagonal elements which correspond to the anomalous Hall
effect. We do not study the relativistic corrections to the
diagonal elements which correspond to the anisotropic
magneto-resistance (AMR) and lead to a difference of order
$1/c^4$ between $\tsig_{xx}$ and $\tsig_{zz}$. Thus, in this
work, the diagonal elements are calculated at order $1/c^0$ and by
consequence are all equal, while the off-diagonal elements are
calculated at order $1/c^2$. To get analytical expressions, we
have made several approximations: free-electron approximation,
weak-scattering limit and weak-relativistic limit for the Dirac
approach. In Sec.~III, we have shown that the conductivity is
equal to
\begin{eqnarray}\label{cond1}
  \tsig_{ij}&=&\frac{e^2\hbar}{4\pi\Omega}
  {\mathrm Tr}\left\langle v_i\left(G^+-G^-\right)v_jG^-\right. \no \\
  & &-\left.v_iG^+v_j\left(G^+-G^-\right)\right\rangle_c \no \\
  & &-\frac{e^2}{4i\pi\Omega}{\mathrm Tr}\left\langle
  \left(G^+-G^-\right)(r_iv_j-r_jv_i)\right\rangle_c,
\end{eqnarray}
where the Green's function $G^{\pm}$ is associated with the total
Hamiltonian: $G^{\pm} = (\epsi_F\pm i0-H)^{-1} = (\epsi_F\pm
i0-H_0-W)^{-1}$ where $H_0$ is the non-perturbed Hamiltonian and
$W$ the perturbation (equal to the potential $V$ in the Dirac
approach and to $V+H_{so}$ in the Pauli approach where $H_{so}$
is the spin-orbit coupling). The explicit form of the potential
$V$ does not enter in the calculations, thus the results obtained
below apply for both impurity scattering and phonon scattering in
the adiabatic approximation. We modelize the compound in the
following way: the total volume of the sample $\Omega=L^3$ is
divided into $N$ cells of volume $\Omega_0=a^3$. In each cell, the
potential takes a constant value $V$ with a probability
distribution $P(V)$ which is characterized by its moments
$\langle V^n\rangle_c = \int P(V)V^ndV$. A proper choice of the
energy origin yields $\langle V\rangle_c=0$. We assume that there
are no correlations in the value of the potential in different
cells. In this first approach, we neglect in (\ref{cond1}) the
contribution of the terms which involves product of two advanced
(or retarded) Green's functions. Such an approximation is
justified in the weak-disorder limit\cite{Rammer98}. In Appendix
B, we have shown that $\tsig^{II}_{ij}$, calculated in the Dirac
approach, contains two parts, the first one related to the orbital
momentum, which is negligible in our model, and the second one
which is exactly compensated by terms in $\tsig^I_{ij}$. Then, we
do not need to calculate this contribution. The conductivity
reduces to
\begin{eqnarray}\label{cond2}
  \tsig_{ij}=\frac{e^2\hbar}{2\pi\Omega}
  {\mathrm Tr}\left\langle v_iG^+v_jG^-\right\rangle_c.
\end{eqnarray}
\wide{m}{
\begin{figure}
\centering \epsfig{file=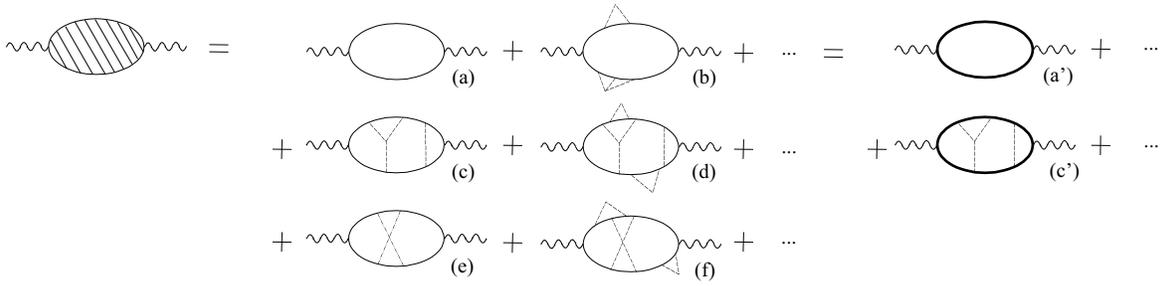} \vspace{0.25cm}
\caption{Illustration of the conductivity with the help of
Feynman diagrams. The total conductivity (hatched diagram),
expressed like an infinite sum of diagrams involving the
non-disordered Green's function $G_0$ (thin curve line), can be
rewrite like an infinite sum of diagrams involving the average
Green's function $\underline{G}$ (bold curve line). The wave
lines refer to the velocity and the dashed lines to the
potential.}\label{figCond}
\end{figure}
}
We introduce first the t-matrix $T=W+WG_0T$ which allows to write
the Green's function as $G=G_0+G_0TG_0$ where $G_0$ is the
non-perturbed Green's function. Inserting this in (\ref{cond2}),
we get
\begin{eqnarray}\label{cond3}
  \tsig_{ij}&=&\frac{e^2\hbar}{2\pi\Omega}{\mathrm Tr}\left\langle v_iG_0^+v_jG_0^-\right\rangle_c \no \\
  & &+\frac{e^2\hbar}{2\pi\Omega}{\mathrm Tr}\left\langle
  v_iG_0^+TG_0^+v_jG_0^-TG_0^-\right\rangle_c.
\end{eqnarray}
This equation can be illustrated with the help of Feynman diagrams
as is done in Fig.~\ref{figCond}. The conductivity $\tsig_{ij}$,
represented by the full diagram, is then express as a sum of an
infinite number of diagrams. Only few of them are depicted in
Fig.~\ref{figCond}: diagram (a) which corresponds to the first
term in (\ref{cond3}) and diagrams from (b) to (f) which are some
representative samples of the kind of diagrams which give the
second term in (\ref{cond3}). The main approximation done in our
calculation is to neglect the crossed diagrams which correspond
to weak-localization corrections, i.e., we neglect diagrams such
(e) and (f) and we keep only the so-called ladder diagrams.
Weak-localization corrections to the anomalous Hall conductivity
are discussed in a separate paper\cite{Dugaev01}. We introduce the
configurational average Green's function $\uG = \langle
G\rangle_c$ which can be written with the help of the self-energy
$\Sigma = \langle WG_0W\rangle_c + \langle WG_0WG_0W\rangle_c +
...$ since $\uG = (\epsi_F-H_0-\Sigma)^{-1}$. When we neglect the
crossed diagrams, (\ref{cond3}) can be written as
\begin{eqnarray}\label{cond4}
  \tsig_{ij}&=&\frac{e^2\hbar}{2\pi\Omega}{\mathrm Tr}\left\langle v_i\uG^+v_j\uG^-\right\rangle_c \no \\
  &&+\frac{e^2\hbar}{2\pi\Omega}{\mathrm Tr}\left\langle
  v_i\uG^+T'\uG^+v_j\uG^-T'\uG^-\right\rangle_c,
\end{eqnarray}
with $T'$ solution of $T'=W+W\uG T'$. The first term in the right
side hand is the so-called bubble term
($\equiv\tsig_{ij}^{bubble}$) and the second one corresponds to
the vertex corrections ($\equiv\tsig_{ij}^{vertex}$). Within this
transformation, the calculation of the conductivity is then
reduced to two distinct problems: determination of the average
Green's function (i.e., the self-energy) and calculation of the
vertex corrections. Because of the weak-scattering limit, we keep
in the self-energy and the t-matrix the lowest sufficient orders
\begin{eqnarray}\label{selfandt}
  \left\{\begin{array}{ll}
  \Sigma=\langle WG_0W\rangle_c \\
  T'=W+W\uG W
  \end{array}\right..
\end{eqnarray}
In the t-matrix, we have to keep the terms up to the second order
with $V$ because it is necessary to go beyond the Born
approximation to get the skew-scattering\cite{Smit}. The explicit
calculation of (\ref{cond4}) in the approximations
(\ref{selfandt}) for the Dirac and Pauli approaches is presented
in the next two sections.

\subsection{Dirac approach}
\begin{figure}
\centering \epsfig{file=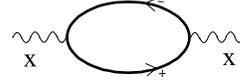}\vspace{0.25cm} \caption{Bubble
diagram contributing to the diagonal conductivity $\tsig_{xx}$.
The signs $+/-$ refer to the retarded/advanced average Green's
function $\underline{G}^{\pm}$.}\label{figB}
\end{figure}

\begin{figure}
\centering \epsfig{file=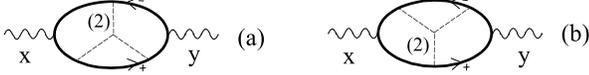}\vspace{0.25cm}
\caption{Diagrams contributing to the off-diagonal conductivity
$\tsig_{xy}$ through the skew-scattering mechanism in both Pauli
and Dirac approaches. The number in bracket indicates the order
with $1/c$ of the matrix elements of the velocity (wave line),
average Green's function (bold curve line) and potential (dashed
line). It is omitted in case of zero order with
$1/c$.}\label{figSS}
\end{figure}

\begin{figure}
\centering \epsfig{file=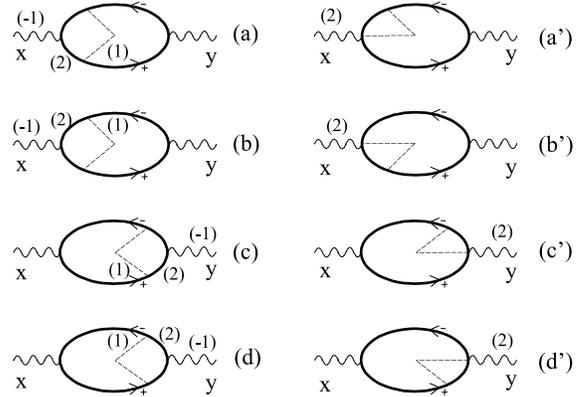}\vspace{0.25cm}
\caption{Diagrams contributing to the off-diagonal conductivity
$\tsig_{xy}$ through the side-jump mechanism in Dirac approach
(left column) and Pauli approach (right column). The number in
bracket indicates the order with $1/c$ of the matrix elements of
the velocity (wave line), average Green's function (bold curve
line) and potential (dashed line).}\label{figSJ}
\end{figure}

We assume free electrons in a uniform effective magnetic field
${\bf B}_{\eff}$ parallel to the z-axis and submitted to a
potential. The non-perturbed part of the Hamiltonian is
\begin{eqnarray}\label{DiracH0}
  H_0=c\left(\balph\cdot\op\right)+(\beta-1)mc^2-\mu_B\beta\sigma_zB_{\eff},
\end{eqnarray}
and the perturbation part is simply the potential $W=V$. The
matrix elements of the average Green's function are
$\bra\uG^{\pm}\ket=\left(\epsi_F-\epsi^s_k\pm
i\hbar/2\tau^s_k\right)^{-1}$ where the eigenvalues $\epsi^s_k$
of (\ref{DiracH0}) are in the weak-relativistic limit equals to
\begin{eqnarray}\label{DiracEignupper}
    \epsi^s_k&=&\frac{\hbar^2k^2}{2m}-s\mu_BB_{\eff}+o\left(\frac{1}{c^2}\right),
\end{eqnarray}
for the upper band, and
\begin{eqnarray}\label{DiracEignlower}
    \epsi^s_{\underline{k}}=-2mc^2+o\left(\frac{1}{c^0}\right),
\end{eqnarray}
for the lower band. The $s$ index refers to the spin ($s=1$ for
spin up and $s=-1$ for spin down), the $k$ index refers to the
upper band and the $\underline{k}$ to the lower band. The
life-time $\tau^s_k$ which appears in the expression of the
average Green's function is given by
\begin{eqnarray}\label{Life-time}
  \frac{\hbar}{2\tau^s_k}&=&-{\mathrm Im}\bra\Sigma^+\ket=-{\mathrm Im}\bra\langle
  VG_0^+V\rangle_c\ket \no \\
  &=&\pi\Omega_0{\cal N}_s(\epsi^s_k)\langle V^2\rangle_c,
\end{eqnarray}
where ${\cal N}_s$ is the density of states of spin $s$ by unit
volume. In the Dirac approach, the velocity ${\bf v}$ is simply
equal to $c\,\balph$. Because we have chosen to work in the basis
where the non-perturbed Hamiltonian $H_0$ (and by consequence the
Green's function $G_0$) is diagonal, we have to calculate the
velocity in this basis, we get
\begin{eqnarray}\label{vbasis}
  {\bf v}=\left(\begin{array}{cc}
  {\bf u}({\bf k})+o\left(\frac{1}{c^2}\right) & c\bsig+o\left(\frac{1}{c}\right) \\
  c\bsig+o\left(\frac{1}{c}\right) & o(c^0)
  \end{array}\right),
\end{eqnarray}
where ${\bf u}({\bf k})$ is the $(2\times 2)$ matrix
\begin{eqnarray}\label{vupper}
  {\bf u}({\bf k})=\frac{\hbar {\bf k}}{m}
  \left(\begin{array}{cc}
  1&0\\
  0&1
  \end{array}\right).
\end{eqnarray}
We have now all the ingredients to calculate the bubble term in
(\ref{cond4})
\begin{eqnarray}\label{barterm}
  \tsig_{ij}^{bubble}&=&\frac{e^2\hbar}{2\pi\Omega}
  \sum_{kss'}\bra v_i\ketsp\brasp\uG^+\ketsp \no \\
  & &\times\brasp v_j\ket\bra\uG^-\ket.
\end{eqnarray}
The configurational average $\langle...\rangle_c$ has been dropped
because in the Dirac approach the velocity is a non-disordered
quantity. At order $1/c^0$, only the diagonal elements ($s=s'$, no
spin-flip) of the velocity and the particles in the upper band
contribute, then we have
\begin{eqnarray}\label{bartermc0}
  \tsig_{ij}^{bubble}=\frac{e^2\hbar^3}{2\pi m^2\Omega}\sum_{ks}
  \frac{k_ik_j}{\left(\epsi_F-\epsi^s_k\right)^2+\frac{\hbar^2}{4\left(\tau^s_k\right)^2}}.
\end{eqnarray}
The dispersion law $\epsi^s_k$ given by (\ref{DiracEignupper}) is
isotropic at order $1/c^0$. Then, the angular dependence is
entirely contained in the factor $k_ik_j$, which means that only
diagonal components of the conductivity are different from zero.
To order $1/c^0$, the vertex corrections to the diagonal
components vanish, so that the total conductivity $\tsig_{ii}$ is
equal to $\tsig_{ii}^{bubble}$. After integration over $k$, we get
\begin{eqnarray}\label{Einrel}
  \tsig_{xx}=e^2{\cal N}_{\uparrow}\frac{l^{\uparrow}v_F^{\uparrow}}{3}
  +e^2{\cal
  N}_{\downarrow}\frac{l^{\downarrow}v_F^{\downarrow}}{3}
  \equiv\tsig_{xx}^{\uparrow}+\tsig_{xx}^{\downarrow},
\end{eqnarray}
which corresponds to the Einstein relation with two spin channels
where $l^s = v_F^s\tau_F^s$ is the mean-free-path, $v_F^s$ and
${\cal N}_s$ are the velocity and the density of states by unit
volume at the Fermi energy for spin s respectively (identical
expressions are obtained for $\tsig_{yy}$ and $\tsig_{zz}$). The
diagram which gives this contribution is depicted on figure
\ref{figB}.
The off-diagonal components of the conductivity arise only when
we take the vertex corrections into account. If we expend the
t-matrix up to the second order in $V$, from (\ref{cond4}), we get
\begin{eqnarray}\label{vertexterm}
  \tsig_{ij}^{vertex}&=&\frac{e^2\hbar}{2\pi\Omega}
  {\mathrm Tr}\left\langle v_i\uG^+(V+V\uG^+V)\uG^+\right. \no \\
  & &\left.\times v_j\uG^-(V+V\uG^-V)\uG^-\right\rangle_c.
\end{eqnarray}
We then need the potential in the new basis
\begin{eqnarray}\label{Potbasis}
  &&V=\tilde{V}({\bf k'}-{\bf k})\left(\begin{array}{cc}
  U({\bf k},{\bf k}')+o\left(\frac{1}{c^4}\right)
  & \frac{\hbar(\bssig\cdot{\bf k'})}{2mc}+o\left(\frac{1}{c^3}\right) \\
  \frac{\hbar(\bssig\cdot{\bf k'})}{2mc}+o\left(\frac{1}{c^3}\right) & o(c^0)
  \end{array}\right),\no\\
\end{eqnarray}
where $U({\bf k},{\bf k}')$ is the $(2\times 2)$ matrix
\wide{m}{
\begin{eqnarray}\label{Potupper}
  U({\bf k},{\bf k}')=\left(\begin{array}{cc}
  1-\frac{\hbar^2(({\bf k'}-{\bf k})^2+2i({\bf k'}\times{\bf k})\cdot{\bf e}_z)}{8m^2c^2}
  & \frac{\hbar^2((k_x'-ik_y')k_z'-(k_x-ik_y)k_z-i({\bf k'}\times{\bf k})\cdot({\bf e}_x-i{\bf e}_y))}{4m^2c^2} \\
  \frac{\hbar^2((k_x'+ik_y')k_z'-(k_x+ik_y)k_z-i({\bf k'}\times{\bf k})\cdot({\bf e}_x+i{\bf e}_y))}{4m^2c^2} &
  1-\frac{\hbar^2(({\bf k'}-{\bf k})^2-2i({\bf k'}\times{\bf k})\cdot{\bf e}_z)}{8m^2c^2}
  \end{array}\right),
\end{eqnarray}
}
and $\tilde{V}({\bf q})=\int d{\bf r}\,e^{i{\bf q}\cdot{\bf
r}}V({\bf r})/\Omega$ is the Fourier transform of the potential.
When we study in detail all the diagrams included in
(\ref{vertexterm}), we see that only two kind of
diagrams\cite{Note2} contribute to the conductivity at order
$1/c^2$. These diagrams are depicted on figures \ref{figSS} and
\ref{figSJ} (left column).

The first series of diagrams (see Fig.~\ref{figSS}) involves
velocities at order $1/c^0$, Green's functions at order $1/c^0$,
which means that only particles in the upper band contribute, and
potential twice at order $1/c^0$ and once at order $1/c^2$ which
ensures a total order of $1/c^2$ for the conductivity. The
diagrams of this first series correspond to the skew-scattering
mechanism. The second series of diagrams (left column in
Fig.~\ref{figSJ}) involves one velocity at order $1/c^0$ and one
at order $c$, three Green's functions at order $1/c^0$ and one at
order $1/c^2$, which means that we have a transition of particles
between the upper and lower bands, and potential one time at
order $1/c^0$ and one time at order $1/c$ which ensures a total
order of $1/c^2$ for the conductivity. The diagrams of this
second series correspond to the side-jump mechanism. In the
following, we present the explicit calculation of the
conductivity due to these two series. Let us start with the
skew-scattering. We present the calculation of the diagram (a) in
Fig.~\ref{figSS}, which gives
\begin{eqnarray}\label{skew1}
  &&\tsig_{xy}^{(\ref{figSS}a)}=\frac{e^2\hbar}{2\pi\Omega}
  \sum_{kk'k''s}\left\langle\bra v_x\ket^{(0)}\bra\uG^+\ket^{(0)}\right. \no \\
  & &\times \bra V\ketkp^{(0)}\brakp\uG^+\ketkp^{(0)}\brakp V\ketkpp^{(0)} \no \\
  & &\times\brakpp\uG^+\ketkpp^{(0)}\brakpp v_y\ketkpp^{(0)}\brakpp\uG^-\ketkpp^{(0)} \no \\
  & &\times\left.\brakpp V\ket^{(2)}\bra\uG^-\ket^{(0)}\right\rangle_c.
\end{eqnarray}
The number in bracket indicates the order with respect to $1/c$ of
the matrix elements like in Fig.~\ref{figSS} when the order is
different from zero. We remark that for a total order $1/c^2$ of
the conductivity, the spin is conserved during the process (no
spin-flip scattering). We insert in this expression, the matrix
elements given by (\ref{vupper}) and (\ref{Potupper}) and perform
the integration over $k$, $k'$ and $k''$. The final contribution
to the conductivity corresponding to the diagram (a) is a complex
quantity. The calculation of the diagram (b) gives the conjugated
expression, then the total contribution due to the
skew-scattering mechanism is a real quantity equal to
\begin{eqnarray}\label{skew2}
  \tsig_{xy}^{SS}
  &=&-\frac{\pi m^2\lambda^2}{6\hbar^2}\frac{\langle V^3\rangle_c}{\langle V^2\rangle_c}
  \left({\cal N}_{\uparrow}\Omega_0\tsig_{xx}^{\uparrow}(v_F^{\uparrow})^2
  -{\cal N}_{\downarrow}\Omega_0\tsig_{xx}^{\downarrow}(v_F^{\downarrow})^2\right) \no \\
  &\equiv&\tsig_{xy}^{SS\uparrow}+\tsig_{xy}^{SS\downarrow}.
\end{eqnarray}
We turn now our attention to the side-jump mechanism. The diagram
(a) of Fig.~\ref{figSJ} gives
\begin{eqnarray}\label{side1}
  &&\tsig_{xy}^{(\ref{figSJ}a)}=\frac{e^2\hbar}{2\pi\Omega}
  \sum_{kk's}\left\langle\bra v_x\ketkbs^{(-1)}\brakbs\uG^+\ketkbs^{(2)}\right. \no \\
  & &\times\brakbs V\ketkp^{(1)}\brakp\uG^+\ketkp^{(0)}\brakp v_y\ketkp^{(0)} \no \\
  & &\left.\times\brakp\uG^-\ketkp^{(0)}\brakp
  V\ket^{(0)}\bra\uG^-\ket^{(0)}\right\rangle_c.\no\\
\end{eqnarray}
In this mechanism, due to the presence of off-diagonal elements
in the velocity (\ref{vbasis}) and the potential
(\ref{Potbasis}), a particle of the upper band $\epsi_k^s$
experiences a virtual transition in the lower band
$\epsi_{\underline{k}}^{-s}$ associated to the opposite spin. We
perform the integrations over $k$ and $k'$, add the contributions
of the four diagrams (\ref{figSJ}a) to (\ref{figSJ}d) and finally
obtained
\begin{eqnarray}\label{side2}
  \tsig_{xy}^{SJ}&=&-e^2{\cal N}_{\uparrow}\frac{2\delta^{\uparrow}v_F^{\uparrow}}{3}
  +e^2{\cal
  N}_{\downarrow}\frac{2\delta^{\downarrow}v_F^{\downarrow}}{3}\no\\
  &\equiv&\tsig_{xy}^{SJ\uparrow}+\tsig_{xy}^{SJ\downarrow},
\end{eqnarray}
where $\delta^s$ is the transverse displacement (or side-jump)
given by $\hbar v_F^s/4mc^2=\lambda^2 k_F^s/4$. The expression of
$\tsig_{xy}$ for the side-jump is similar to the expression
(\ref{Einrel}) of $\tsig_{xx}$ but instead of the mean-free-path
$l$, we have $2\delta$. In contrast to $\tsig_{xy}^{SS}$, the
side-jump contribution to the off-diagonal conductivity is
independent of disorder.

In the case of a parabolic band, the Einstein relation
(\ref{Einrel}) reduces to the Drude formula with two spin channels
\begin{eqnarray}\label{Drude}
  \tsig_{xx}=e^2\frac{n_{\uparrow}\tau_F^\uparrow}{m}+e^2\frac{n_{\downarrow}\tau_F^\downarrow}{m},
\end{eqnarray}
where $n_s=m{\cal N}_s(v_F^s)^2/3$ is the electron density for
spin s. The skew-scattering (\ref{skew2}) and side-jump
(\ref{side2}) contributions yield
\begin{eqnarray}\label{skewDrude}
  \tsig_{xy}^{SS}=-\frac{e^2\lambda^2}{2\hbar}\frac{\pi\Omega_0\langle V^3\rangle_c}{\hbar\langle V^2\rangle_c}
  (n_{\uparrow}^2\tau_F^\uparrow-n_{\downarrow}^2\tau_F^\downarrow),
\end{eqnarray}
and,
\begin{eqnarray}\label{sideDrude}
  \tsig_{xy}^{SJ}=-\frac{e^2\lambda^2}{2\hbar}(n_{\uparrow}-n_{\downarrow}).
\end{eqnarray}

\subsection{Pauli approach}
In the Pauli approach, the Hamiltonian given by (\ref{PauliH0})
is the sum of a non-perturbed part and a perturbation $W$ given
by (\ref{PauliW}) which contains the potential and the spin-orbit
coupling. The velocity associated with this Hamiltonien consists
of a normal part and an anomalous part due to the spin-orbit
coupling
\begin{eqnarray}\label{Pauliv}
  {\bf v}=\frac\op{m}+\frac{\hbar}{4m^2c^2}\left(\bsig\times\bnab V\right).
\end{eqnarray}
We neglect the contribution of the spin-orbit coupling in the
life-time, then the average Green's function is $\uG^{\pm} =
(\epsi_F-H_0\mp i\hbar/\tau_k^s)^{-1}$ where $\tau_k^s$ is given
by (\ref{Life-time}). Because of this approximation, the
derivation of the diagonal conductivity $\tsig_{xx}$ is similar
to the one done in the Dirac approach and we obtain the expression
(\ref{Einrel}).

The off-diagonal elements of the conductivity are obtained from
the vertex corrections. For the skew-scattering, the diagrams
which contribute are exactly the same than in the Dirac approach
(see Fig.~\ref{figSS}) because the only matrix elements of the
potential (\ref{Potupper}) which contribute in the Dirac approach
correspond precisely to the matrix elements of the potential in
the Pauli approach (see Eq. (\ref{mateleV})). The other terms in
(\ref{Potupper}) are Darwin-like terms and do not contribute to
the off-diagonal conductivity. Then, in the Pauli approach, the
skew-scattering mechanism corresponds to the same Feynman
diagrams and gives the same final expression (\ref{skew2}) as the
weak-relativistic limit of the Dirac approach.

Concerning the side-jump, the correspondence between the two
approaches is not so simple. In the Dirac approach, we have seen
that a virtual transition occurs from the upper band to the lower
band. In the Pauli approach, no such a transition can take place
because there is only one band. However, we have a supplementary
part in the velocity, the anomalous velocity which is of order
$1/c^2$ and leads to the side-jump mechanism. The corresponding
diagrams are depicted on the right column of Fig.~\ref{figSJ}.
For each diagram in the left column (i.e., in the Dirac approach),
we have an equivalent diagram in the right column (i.e., in the
Pauli approach). The change between the left and right column
corresponds to a vertex renormalization because the matrix
elements of the product ${\bf v}GV$ in the Dirac approach are
equal to the matrix elements of ${\bf v}$ in the Pauli approach
\begin{eqnarray}\label{Paulivel}
  \langle k,s|{\bf v}|k',s'\rangle&=&\frac{\hbar{\bf
  k}}{m}\delta_{kk'}\delta_{ss'}\no\\
  &+&\tilde{V}({\bf k'}-{\bf k})\frac{i\hbar}{4m^2c^2}\bsig_{ss'}\times\left({\bf k}-{\bf
  k'}\right).
\end{eqnarray}
Thus, when we calculate, for example the diagram (\ref{figSJ}a')
\begin{eqnarray}\label{side3}
  \tsig_{xy}^{(\ref{figSJ}a')}&=&\frac{e^2\hbar}{2\pi\Omega}
  \sum_{kk's}\left\langle\bra v_x\ketkp^{(2)}\brakp\uG^+\ketkp^{(0)}\right. \no \\
  & &\times\brakp v_y\ketkp^{(0)}\brakp\uG^-\ketkp^{(0)} \no \\
  & &\times\left.\brakp V\ket^{(0)}\bra\uG^-\ket^{(0)}\right\rangle_c,
\end{eqnarray}
we obtain the same contribution than from the expression
(\ref{side1}) of the diagram (\ref{figSJ}a). The final result,
after summation over the four diagrams (\ref{figSJ}a') to
(\ref{figSJ}d'), is then identical to (\ref{side2}).

\section{Discussion}
We now briefly discuss the influence of impurity scattering and
phonon scattering on the resistivity and on the anomalous Hall
resistivity, which are, in the limit $\tsig_{xy}\ll\tsig_{xx}$
simply given by $\tilde{\rho}_{xx}\simeq 1/\tsig_{xx}$ and
$\tilde{\rho}_{H}=-\tilde{\rho}_{xy}\simeq
\tsig_{xy}/\tsig_{xx}^2$. The only terms which depend on the
scattering in the expressions of $\tsig_{xy}$ given by
(\ref{skew2}) and (\ref{side2}) and of $\tsig_{xx}$ given by
(\ref{Einrel}), are the moments $\langle V^2\rangle_c$ and
$\langle V^3\rangle_c$. Indeed, we have
\begin{eqnarray}
    \left\{\begin{array}{lll}
  \tsig_{xx}\propto \frac{1}{\langle V^2\rangle_c} \\
  \tsig_{xy}^{SS}\propto \frac{\langle V^3\rangle_c}{\langle V^2\rangle_c^2} \\
  \tsig_{xy}^{SJ}\,\,{\mathrm indep.\,\,of\,\,}\langle V^n\rangle_c
  \end{array}\right..
\end{eqnarray}
Then, the variations with the moments $\langle V^2\rangle_c$ and
$\langle V^3\rangle_c$ of the resistivity and the anomalous Hall
resistivity are like $\tilde{\rho}_{xx}\propto \langle
V^2\rangle_c$, $\tilde{\rho}_{xy}^{SS}\propto \langle
V^3\rangle_c$ and $\tilde{\rho}_{xy}^{SJ}\propto \langle
V^2\rangle_c^2$.

To illustrate the dependence with disorder in the case of
impurity scattering, we consider a binary alloy $A_xB_{1-x}$ for
which
\begin{eqnarray}\label{potmoy2}
  \langle V^2\rangle_c=x(1-x)(\epsi_A-\epsi_B)^2,
\end{eqnarray}
and,
\begin{eqnarray}\label{potmoy3}
  \langle V^3\rangle_c=x(1-x)(1-2x)(\epsi_A-\epsi_B)^3,
\end{eqnarray}
where $\epsi_{A(B)}$ is the value of the potential $V$ on site
$A(B$) and $x$ is the concentration of sites $A$. Keeping the
lowest orders in $x$ (weak-disorder limit), we get
\begin{eqnarray}\label{resalloy}
  \left\{\begin{array}{lll}
  \tilde{\rho}_{xx}\propto x \\
  \tilde{\rho}_{xy}^{SS}\propto (x-3x^2) \\
  \tilde{\rho}_{xy}^{SJ}\propto x^2
  \end{array}\right.,
\end{eqnarray}
which is in agreement with the simple relation given by
(\ref{anomres}) but in contradiction with the common belief that
the quadratic term would arise only from the side-jump mechanism.
In fact, the skew-scattering mechanism, which is responsible for
the linear term, gives also an important contribution to the
quadratic term, a result that Kondorskii {\it et al.}
\cite{Kondorskii} have already obtained. In addition, our
calculations specify all the approximations which founded the
relation (\ref{anomres}) and show that it should not be valid in
the general case, in particular for high-disordered system,
high-relativistic limit, complex band structure or in the case of
heterogeneous systems, such as thin films or multilayers.

Something may also be said about phonon scattering. Due to the
fluctuating sign of the potential generated by phonons, the third
moment $\langle V\rangle^3_c$ can be expected to be very small
and accordingly, the skew-scattering contribution (\ref{skew2})
to the conductivity is negligible\cite{Chien80a}. We have then
\begin{eqnarray}\label{resphonon}
  \left\{\begin{array}{lll}
  \tilde{\rho}_{xx}\propto \langle V^2\rangle_c \\
  \tilde{\rho}_{xy}^{SS}\simeq 0 \\
  \tilde{\rho}_{xy}^{SJ}\propto \langle V^2\rangle^2_c
  \end{array}\right.,
\end{eqnarray}
which yields the simple relation
$\tilde{\rho}_{xy}\propto\tilde{\rho}_{xx}^2$, in agreement with
experimental results\cite{Chien80b}.

The Hall angle, which corresponds to the angle between the
electric field and the charge current, is an important quantity.
For an applied electric field in the x-direction and an effective
magnetic field in the z-direction, we have tg$(\theta_H)\equiv
j_y/j_x=\tsig_{yx}/\tsig_{xx}$. The conductivity elements
$\tsig_{xx}$ and $\tsig_{yx}$ are in a first approximation the
sums of contributions due to spins up and down. We can thus
define a spin-dependent Hall angle
\begin{eqnarray}\label{HallAngleDef}
  {\mathrm tg}(\theta_H^{\uparrow(\downarrow)})\equiv
  \frac{j_y^{\uparrow(\downarrow)}}{j_x^{\uparrow(\downarrow)}}
  =\frac{\tsig_{yx}^{\uparrow(\downarrow)}}{\tsig_{xx}^{\uparrow(\downarrow)}}.
\end{eqnarray}
We  insert the expressions (\ref{skew2}) and (\ref{side2}) of the
skew-scattering and side-jump off-diagonal conductivities as well
as the expression (\ref{Einrel}) of the diagonal conductivity and
obtain for spin s
\begin{eqnarray}\label{HallAngle}
  \theta_H^s\approx s\left(\frac{2\delta^s}{l^s}+\frac{\pi m^2\lambda^2}{6\hbar^2}
  \frac{\langle V^3\rangle_c}{\langle V^2\rangle_c}
  {\cal N}_s\Omega_0(v_F^s)^2\right).
\end{eqnarray}
$\theta_H^{\uparrow}$ and $\theta_H^{\downarrow}$ are not only
opposite in sign, they take distinct absolute values due to spin
polarization. As a consequence, the spin current $({\bf
j}^{\uparrow}-{\bf j}^{\downarrow})$ has a longitudinal (i.e.,
along the x-axis) and a transverse (i.e., along the y-axis)
component and the charge current $({\bf j}^{\uparrow}+{\bf
j}^{\downarrow})$ acquires a transverse component which
corresponds to the anomalous Hall effect. In a paramagnetic
material, Eq. (\ref{HallAngle}) yields
$\theta_H^{\uparrow}=-\theta_H^{\downarrow}$ and both the
transverse component of the charge current and the longitudinal
component of the spin current vanish. However, the transverse
component of the spin current remains. It corresponds precisely to
the ``spin Hall effect'' recently proposed\cite{HirschZhang}.

In the case of impurity scattering and in the weak-disorder limit,
the magnitude of the Hall angle is determined mostly by the
skew-scattering contribution. Indeed, in this limit we have
$\theta_H^{SJ}\approx 2\delta/l\approx 10^{-3}$~rad whereas
$\theta_H^{SS}\approx\pi(1-2x)(\epsi_A-\epsi_B)\epsi_F/3mc^2W\approx
5.10^{-2}$~rad where we have taken $l\approx 200$~\AA, $x\approx
0.2$, $\epsi_A-\epsi_B\approx 2$~eV, $\epsi_F\approx 10$~eV, the
band width $W\approx5$~eV, $mc^2\simeq500$~keV and a band factor
$\alpha\approx10^4$. For simplicity, we have dropped the spin
index. This order of magnitude is consistent with experimental
results\cite{Chien80a}. When the disorder increases, the
mean-free-path $l$ decreases significantly which means, since the
quantity $\delta$ is disorder independent, an increase in the
side-jump contribution to the Hall angle. However, the
skew-scattering contribution to the Hall angle increases in the
same way. It is thus not possible to predict in this first
approach which contribution is dominant in the high-disorder
regime.

In the case of phonon scattering, the Hall angle contains mostly
the side-jump contribution $\theta_H=\theta_H^{SJ}\approx
2\delta/l$ which is of order $\approx10^{-2}$~rad where we have
used $\delta\approx10^{-11}$~m and $l=\tau v_F$ with a relaxation
time $\tau\approx 10^{-15}$~s and $v_F \approx 10^6$~m.s$^{-1}$.

To summarize, we have, in this article, proposed a model based on
the Dirac equation and on the Kubo formalism which allows to
calculate on the same footing the anomalous Hall conductivity due
to both skew-scattering and side-jump mechanisms. The consistency
of this approach with the one based on the Pauli equation has been
studied in detail in the weak-relativistic limit. In particular,
we have shown that in order to calculate the anomalous Hall
conductivity one has to consider in the Dirac approach the total
conductivity $\tsig^I_{ij}+\tsig^{II}_{ij}$, otherwise unphysical
results are obtained. Next, we applied our model to treat a
disordered ferromagnetic bulk compound submitted to a potential in
the free electron approximation, weak-scattering and
weak-relativistic limits. By these means, we have obtained
explicit expressions for the anomalous Hall conductivity for both
skew-scattering and side-jump mechanisms (given by (\ref{skew2})
and (\ref{side2})). In addition, we have highlighted the
difference concerning the Feynman diagrams describing the
side-jump mechanism in the Dirac and Pauli approaches and have
shown that it corresponds to different vertex renormalizations.
\end{multicols}
\appendix
\begin{multicols}{2}
\section{From the Kubo formula to the Streda formula}
In the linear response approximation, Kubo has shown that the
conductivity tensor is related to a two currents correlation
function\cite{Kubo}
\begin{eqnarray}\label{kubo1}
  \tsig_{ij}(\omega)=\Omega\lims\int_0^{\beta} d\lambda\int_0^{+\infty} dt\,
  e^{\frac{it}{\hbar}(-\hbar\omega+is)} \no \\
  \times {\mathrm Tr}\bigg\langle\rho_0J_j(0)J_i(t+i\hbar\lambda)\bigg\rangle_c,
\end{eqnarray}
where it is assumed that the applied field leads to a
time-dependent perturbation of the form: $H'(t)=H'_0{\mathrm
exp}(\frac{it}{\hbar}(-\hbar\omega+is))$. $\Omega$ is the volume
of the sample, $\beta\equiv1/k_BT$, $\rho_0$ is the density
matrix in equilibrium in absence of perturbation, $J_i$ is the
i-component of the current density operator in the Heisenberg
representation and $\langle...\rangle_c$ denotes the
configurational average. Following Luttinger\cite{Luttinger69}, we
obtain in the independent electrons approximation
\begin{eqnarray}\label{current}
  \langle n|J_i(t+i\hbar\lambda)|m\rangle=e^{\frac{i}{\hbar}(t+i\hbar\lambda)(\epsi_n-\epsi_m)}\langle
  n|\tJ_i|m\rangle,
\end{eqnarray}
where we have used $H=\sum_n\epsi_na_n^+a_n$ and defined $\tJ$ as
the current density operator in the Schr\"odinger representation.
Using the relation
Tr$\left[\rho_0a_m^+a_na_p^+a_q\right]=\delta_{mq}\delta_{np}f(\epsi_m)(1-f(\epsi_n))$
where $f(\epsi)$ is the Fermi-Dirac distribution function, we get
\wide{m}{
\begin{eqnarray}\label{kubo2}
  \tsig_{ij}(\omega)=\Omega\lims\int_0^{\beta} d\lambda\,\,e^{-\lambda(\epsi_n-\epsi_m)}
  \int_0^{+\infty} dt\sum_{nm}\bigg\langle f(\epsi_m)(1-f(\epsi_n))
  e^{\frac{it}{\hbar}(-\hbar\omega+is+\epsi_n-\epsi_m)}
  \langle m|\tJ_j|n\rangle\langle n|\tJ_i|m\rangle\bigg\rangle_c.
\end{eqnarray}
The integration over $\lambda$ leads to a factor
$(1-e^{-\beta(\epsi_n-\epsi_m)})/(\epsi_n-\epsi_m)$ which can be
simplified with $f(\epsi_m)\left(1-f(\epsi_n)\right)$ as
\begin{eqnarray}\label{functdist}
 \frac{1-e^{-\beta(\epsi_n-\epsi_m)}}{\epsi_n-\epsi_m}f(\epsi_m)\left(1-f(\epsi_n)\right)
 =\frac{f(\epsi_m)-f(\epsi_n)}{\epsi_n-\epsi_m}.
\end{eqnarray}
Inserting this in (\ref{kubo2}) and performing the integration
over $t$, we obtain
\begin{eqnarray}\label{kubo4}
  \tsig_{ij}(\omega)=i\hbar\,\Omega
  \lims\sum_{nm}\left\langle\frac{f(\epsi_m)-f(\epsi_n)}{(\epsi_n-\epsi_m)(\epsi_n-\epsi_m-\hbar\omega+is)}
  \langle m|\tJ_j|n\rangle\langle n|\tJ_i|m\rangle\right\rangle_c.
\end{eqnarray}
}
We shall now make some transformations of this expression in order
to get the Bastin formula.

We restrict our derivation to zero frequency (from now, we drop
the $\omega$ variable). After inserting the identity
$\int_{-\infty}^{+\infty}d\epsi\,\delta(\epsi-H)=1$ in
(\ref{kubo4}), we obtain
\begin{eqnarray}\label{bastin1}
  & &\tsig_{ij}=i\hbar\,\Omega
  \lims\int_{-\infty}^{+\infty}d\epsi\sum_{nm}\left\langle\left(\frac{f(\epsi)\delta(\epsi-\epsi_m)}{(\epsi_n-\epsi)(\epsi_n-\epsi+is)}
  \right.\right.\no \\
  & &-\left.\left.\frac{f(\epsi)\delta(\epsi-\epsi_n)}{(\epsi-\epsi_m)(\epsi-\epsi_m+is)}\right)
  \langle m|\tJ_j|n\rangle\langle n|\tJ_i|m\rangle\right\rangle_c.
\end{eqnarray}
We remark that
\begin{eqnarray}\label{derivee}
&&\lims\frac{1}{(\epsi_n-\epsi)(\epsi_n-\epsi+is)}=\lims\frac{d}{d\epsi}\left(\frac{1}{\epsi_n-\epsi+is}\right);\no
\\
&&
\end{eqnarray}
then we have
\begin{eqnarray}\label{bastin2}
  & &\tsig_{ij}=-i\hbar\,\Omega\lims\int_{-\infty}^{+\infty}d\epsi f(\epsi) \no \\
  & &\times\sum_{nm}\left\langle\langle m|\tJ_j|n\rangle\frac{d}{d\epsi}\left(\frac{1}{\epsi-\epsi_n-is}\right)
  \langle n|\tJ_i|m\rangle\delta(\epsi-\epsi_m) \right. \no \\
  & &\left.-\langle m|\tJ_j|n\rangle\delta(\epsi-\epsi_n)
  \langle n|\tJ_i|m\rangle\frac{d}{d\epsi}\left(\frac{1}{\epsi-\epsi_m+is}\right)\right\rangle_c,
\end{eqnarray}
which can be expressed as
\begin{eqnarray}\label{bastin3}
  \tsig_{ij}=\frac{ie^2\hbar}{\Omega}\int_{-\infty}^{+\infty}d\epsi f(\epsi)
  {\mathrm Tr}\left\langle v_i\frac{dG^+(\epsi)}{d\epsi}v_j\delta(\epsi-H)\right. \no \\
  -\left.v_i\delta(\epsi-H)v_j\frac{dG^-(\epsi)}{d\epsi}\right\rangle_c,
\end{eqnarray}
where we have introduced the Green's function $G^{\pm}(\epsi) =
\lims (\epsi-H\pm is)^{-1}$ and the velocity through the relation
$\tilde{\bf J}=-e{\bf v}/\Omega$. This expression for the
conductivity was first obtained by Bastin {\it et
al.}\cite{Bastin71} but in the particular case of a Schr\"odinger
Hamiltonian and made explicit use of the form taken by the
velocity operator in the Schr\"odinger case. The present
derivation is more general in the sense that it is independent of
the explicit form of the velocity operator and is therefore valid
both for the Schr\"odinger, Pauli and Dirac cases. The only
restriction is the independent electrons approximation. This
formula, called Bastin formula, is interesting because it
expresses the conductivity as a product of velocities and Green's
functions. However, it is still difficult to calculate because of
the integration over the energy $\epsi$. By making an integration
by parts, a factor $df(\epsi)/d\epsi$ appears instead of the
factor $f(\epsi)$ and the integration interval will be thus
reduced.

In (\ref{bastin3}), we express the delta function in terms of
Green's functions using $\delta(\epsi-H) =
-(G^+(\epsi)-G^-(\epsi)) / 2i\pi$. We keep one half of this
expression and make an integration by parts on the second half,
then we get
\wide{m}{
\begin{eqnarray}\label{streda1}
  & &\tsig_{ij}=-\frac{e^2\hbar}{4\pi\Omega}\int_{-\infty}^{+\infty}d\epsi\frac{d f(\epsi)}{d\epsi}
  {\mathrm Tr}\bigg\langle v_i\left(G^+(\epsi)-G^-(\epsi)\right)v_jG^-(\epsi)
  -v_iG^+(\epsi)v_j\left(G^+(\epsi)-G^-(\epsi)\right)\bigg\rangle_c \no \\
  & &+\frac{e^2\hbar}{4\pi\Omega}\int_{-\infty}^{+\infty}d\epsi f(\epsi)
  {\mathrm Tr}\left\langle v_i\frac{dG^-(\epsi)}{d\epsi}v_jG^-(\epsi)-v_iG^-(\epsi)v_j\frac{dG^-(\epsi)}{d\epsi}
  +v_iG^+(\epsi)v_j\frac{dG^+(\epsi)}{d\epsi}-v_i\frac{dG^+(\epsi)}{d\epsi}v_jG^+(\epsi)\right\rangle_c.
\end{eqnarray}
}
The second term in this expression can be simplified by using the
relations $dG^{\pm} (\epsi) / d\epsi = - (G^{\pm}(\epsi))^2$ and
$i\hbar v_i = [r_i,H] = -[r_i,G^{-1}]$ and by performing once more
an integration by parts. Finally, the conductivity can be written
as a sum of two terms $\tsig_{ij}=\tsig_{ij}^{I}+\tsig_{ij}^{II}$
where
\begin{eqnarray}\label{streda2}
  \tsig_{ij}^I&=&-\frac{e^2\hbar}{4\pi\Omega}\int_{-\infty}^{+\infty}d\epsi\frac{d f(\epsi)}{d\epsi} \no \\
  & &\times {\mathrm Tr}\bigg\langle v_i\left(G^+(\epsi)-G^-(\epsi)\right)v_jG^-(\epsi) \no \\
  & &-v_iG^+(\epsi)v_j\left(G^+(\epsi)-G^-(\epsi)\right)\bigg\rangle_c,
\end{eqnarray}
and
\begin{eqnarray}\label{streda3}
  \tsig_{ij}^{II}&=&
  \frac{e^2}{4i\pi\Omega}\int_{-\infty}^{+\infty}d\epsi \frac{df(\epsi)}{d\epsi} \no \\
  & &\times {\mathrm Tr}\bigg\langle\left(G^+(\epsi)-G^-(\epsi)\right)(r_iv_j-r_jv_i)\bigg\rangle_c.
\end{eqnarray}
(\ref{streda2}) and (\ref{streda3}) correspond to the formula
obtained by Streda\cite{Streda82} in the Schr\"odinger case. The
present derivation shows that it holds also in the Pauli and
Dirac cases. For the diagonal components of the conductivity
tensor, $\tsig_{ij}^{II}$ is equal to zero and we obtain the
Kubo-Greenwood formula\cite{Greenwood58}
\begin{eqnarray}\label{greenwood}
  \tsig_{ii}=\frac{e^2\hbar}{4\pi\Omega}\int_{-\infty}^{+\infty}d\epsi\frac{d f(\epsi)}{d\epsi}{\mathrm Tr}\bigg\langle
  v_i\left(G^+(\epsi)-G^-(\epsi)\right) \no \\
  \times v_i\left(G^+(\epsi)-G^-(\epsi)\right)\bigg\rangle_c.
\end{eqnarray}
At zero temperature, the factor $d f(\epsi)/d\epsi$ is equal to
$-\delta(\epsi-\epsi_F)$, only electrons at the Fermi level
contribute to the conductivity (for both diagonal and
off-diagonal components). In conclusion, at $\omega=0$ and $T=0$,
the conductivity tensor can be expressed as a sum of two terms
$\tsig_{ij}=\tsig_{ij}^I+\tsig_{ij}^{II}$ with
\begin{eqnarray}\label{streda4}
  \tsig_{ij}^I=\frac{e^2\hbar}{4\pi\Omega}
  {\mathrm Tr}\bigg\langle v_i\left(G^+-G^-\right)v_jG^- \no \\
  -v_iG^+v_j\left(G^+-G^-\right)\bigg\rangle_c,
\end{eqnarray}
and,
\begin{eqnarray}\label{streda5}
  \tsig_{ij}^{II}=-\frac{e^2}{4i\pi\Omega}{\mathrm Tr}\bigg\langle\left(G^+-G^-\right)(r_iv_j-r_jv_i)\bigg\rangle_c,
\end{eqnarray}
where we have dropped the energy reference $\epsi_F$ by
introducing the Green's functions at the Fermi level
$G^{\pm}=G(\epsi_F\pm i0)=\left(\epsi_F\pm i0-H\right)^{-1}$.

\section{Streda formula in the weak-relativistic limit}
In this appendix, we give the detail of the calculation
concerning the weak-relativistic expansion of the Streda
conductivity starting from the Dirac equation. From
(\ref{streda4}), we see that $\tsig_{ij}^I$ is a combination of
terms such as
\begin{eqnarray}\label{lambda1}
  \Lambda_{ij}(z_1,z_2)=\frac{e^2\hbar}{4\pi\Omega}{\mathrm Tr}\bigg\langle v_iG(z_1)v_jG(z_2)\bigg\rangle_c,
\end{eqnarray}
where $z_1$ and $z_2$ are equals to $\epsi_F\pm i0$. When we
insert the Dirac velocity (\ref{Diracvelo}) and the Dirac Green's
function (\ref{Diracgreen}) in (\ref{lambda1}), make the explicit
product of the 4 operators and take the trace over the lower and
upper components of the wave function, we obtain the general form:
\wide{m}{
\begin{eqnarray}\label{lambda2}
  &&\Lambda_{ij}(z_1,z_2)=
  \frac{e^2\hbar}{4\pi\Omega}{\mathrm Tr}\bigg\langle\sigma_iD(z_1)\frac{\bsig\cdot{\bf
  p}}{2m}\tG(z_1)\sigma_jD(z_2)\frac{\bsig\cdot\op}{2m}\tG(z_2) \no \\
  &&+\sigma_i\tG(z_1)\frac{\bsig\cdot{\bf
  p}}{2m}Q^{-1}(z_1)D(z_1)Q(z_1)\sigma_j\tG(z_2)\frac{\bsig\cdot\op}{2m}Q^{-1}(z_2)D(z_2)Q(z_2) \no \\
  &&+\sigma_iD(z_1)\left(\frac{1}{2m}+\frac{\bsig\cdot{\bf
  p}}{2m}\tG(z_1)\frac{\bsig\cdot\op}{2m}\right)
  \sigma_j\left(\tG(z_2)-\tG(z_2)\frac{\bsig\cdot{\bf
  p}}{2mc}(z_2-V-\mu_B(\bsig\cdot{\bf B}\eff))D(z_2)\frac{\bsig\cdot\op}{2mc}\tG(z_2)\right) \no \\
  &&+\sigma_i\left(\tG(z_1)-\tG(z_1)\frac{\bsig\cdot{\bf
  p}}{2mc}(z_1-V-\mu_B(\bsig\cdot{\bf B}\eff))D(z_1)\frac{\bsig\cdot\op}{2mc}\tG(z_1)\right)
  \sigma_jD(z_2)\left(\frac{1}{2m}+\frac{\bsig\cdot{\bf
  p}}{2m}\tG(z_2)\frac{\bsig\cdot\op}{2m}\right)\bigg\rangle_c.
\end{eqnarray}
Similarly, when we insert (\ref{Diracvelo}) and (\ref{Diracgreen})
in (\ref{streda5}), we get
\begin{eqnarray}\label{sigmaII2}
  \tsig_{ij}^{II}=-\frac{e^2}{4i\pi\Omega}{\mathrm Tr}\bigg\langle\left(\tG^+\frac{\bsig\cdot{\bf
  p}}{2m}(Q^+)^{-1}D^+Q^++D^+\frac{\bsig\cdot\op}{2m}\tG^+
  -\tG^-\frac{\bsig\cdot\op}{2m}(Q^-)^{-1}D^-Q^--D^-\frac{\bsig\cdot{\bf
  p}}{2m}\tG^-\right)\left(r_i\sigma_j-r_j\sigma_i\right)\bigg\rangle_c.
\end{eqnarray}
}
Expressions (\ref{lambda2}) and (\ref{sigmaII2}) are exact
expressions without any assumption on the value of $c$. We will
now calculate the weak-relativistic expansion of these expressions
at orders $1/c^0$ and $1/c^2$ in order to compare them with the
expression obtained from the Pauli approach.

\subsection{Dirac conductivity at order $1/c^0$}
In the non-relativistic limit, $D$ is simply equal to the unit
matrix (see (\ref{expD})), thus (\ref{lambda2}) can be rewritten
as
\begin{eqnarray}\label{lambda4}
  \Lambda_{ij}^{(0)}(z_1,z_2)&=&\frac{e^2\hbar}{4\pi\Omega}
  {\mathrm Tr}\left\langle \frac{p_i}{m}\tG(z_1)\frac{p_j}{m}\tG(z_2)\right. \no \\
  &&+\left.\frac{1}{2m}\left(\sigma_i\sigma_j\tG(z_2)+\sigma_j\sigma_i\tG(z_1)\right)\right\rangle_c,
\end{eqnarray}
where we have used the fact that
$\sigma_i\left(\bsig\cdot\op\right)+\left(\bsig\cdot\op\right)\sigma_i=2p_i$.
When we insert this expression in (\ref{streda4}), the
conductivity $\tsig_{ij}^I$ at order $1/c^0$ is
\begin{eqnarray}\label{sigmaI2}
  \tsig_{ij}^{I(0)}&=&\frac{e^2\hbar}{4\pi\Omega}
  {\mathrm Tr}\left\langle \frac{p_i}{m}(\tG^+-\tG^-)\frac{p_j}{m}\tG^-\right. \no \\
  & &\left.-\frac{p_i}{m}\tG^+\frac{p_j}{m}(\tG^+-\tG^-)\right\rangle_c \no \\
  & &+\varepsilon_{ijk}\frac{e^2\hbar}{4i\pi m\Omega}
  {\mathrm Tr}\left\langle\sigma_k\left(\tG^+-\tG^-\right)\right\rangle_c,
\end{eqnarray}
where $\epsi_{ijk}=1$ if $\{i,j,k\}=\{x,y,z\}$ or cyclic
permutations and $\epsi_{ijk}=0$ otherwise. This factor is
introduce through the term
$(\sigma_i\sigma_j-\sigma_j\sigma_i)=2i\epsi_{ijk}\sigma_k$. The
first term on the right hand side corresponds exactly to the
contribution that we get in a non-relativistic description
because in this case the velocity is ${\bf \tilde{v}}=\op/m$ and
the Green's function is simply the non-relativistic Green's
function $\tG$. In contrast, the second term is not present in the
Pauli approach and should not appear when we take the
non-relativistic limit in the Dirac approach. In fact, we show
below that this term is exactly cancelled by an opposite term in
$\tsig_{ij}^{II(0)}$. Replacing $D$ by $1$ in (\ref{sigmaII2}),
we get ${\tilde\sigma}_{ij}^{II}$ at order $1/c^0$
\begin{eqnarray}\label{sigmaII4}
  {\tilde\sigma}_{ij}^{II(0)}=-\frac{e^2}{4i\pi\Omega}{\mathrm Tr}\left\langle\left(\tG^+-\tG^-\right)
  \left(r_i\frac{p_j}{m}-r_j\frac{p_i}{m}\right)\right\rangle_c \no \\
  -\varepsilon_{ijk}\frac{e^2\hbar}{4i\pi m\Omega}{\mathrm Tr}\left\langle\sigma_k
  \left(\tG^+-\tG^-\right)\right\rangle_c.
\end{eqnarray}
where we have used the relations $\left(\bsig\cdot{\bf
A}\right)\left(\bsig\cdot{\bf B}\right)=\left({\bf A}\cdot{\bf
B}\right)+i\bsig\cdot\left({\bf A}\times{\bf B}\right)$ and
$[r_i,p_j]=i\hbar\delta_{ij}$. The second term in the right hand
side cancels the supplementary term in (\ref{sigmaI2}) and we
obtain finally for the total Dirac conductivity at order $1/c^0$
\begin{eqnarray}\label{totcond0}
  & &\tsig_{ij}^{(0)}=\tsig_{ij}^{I(0)}+\tsig_{ij}^{II(0)} \no \\
  & &=\frac{e^2\hbar}{4\pi\Omega}{\mathrm Tr}\left\langle \frac{p_i}{m}(\tG^+-\tG^-)\frac{p_j}{m}\tG^-
  -\frac{p_i}{m}\tG^+\frac{p_j}{m}(\tG^+-\tG^-)\right\rangle_c \no \\
  & &-\frac{e^2}{4i\pi\Omega}{\mathrm Tr}\left\langle\left(\tG^+-\tG^-\right)
  \left(r_i\frac{p_j}{m}-r_j\frac{p_i}{m}\right)\right\rangle_c,
\end{eqnarray}
which corresponds exactly to the total conductivity obtained from
(\ref{streda4}) and (\ref{streda5}) when we insert the
non-relativistic Pauli velocity ${\bf \tilde{v}}=\op/m$ and the
non-relativistic Pauli Green's function $\tG$.

\subsection{Dirac conductivity at order $1/c^2$}
To get $\tsig_{ij}$ at order $1/c^2$, it is necessary to take
into account the next terms in the expansion of $D$ given by
(\ref{expD}): $D(z)\approx1-Q(z)(z-V-\mu_B(\bsig\cdot{\bf
B}_{\eff}))/2mc^2$. Thus, from (\ref{lambda2}), we get
\wide{m}{
\begin{eqnarray}\label{lambda5}
  &&\Lambda_{ij}^{(2)}(z_1,z_2)=
  \frac{e^2\hbar}{4\pi\Omega}
  {\mathrm Tr}\bigg\langle\frac{p_i}{m}\tG(z_1)({\bf v}_{rc})_j\tG(z_2)+({\bf v}_{rc})_j\tG(z_1)\frac{p_j}{m}\tG(z_2) \no \\
  & &+\frac{p_i}{m}\tG(z_1)\frac{p_j}{m}\tG(z_2)H_{rc}\tG(z_2)
  +\frac{p_i}{m}\tG(z_1)H_{rc}\tG(z_1)\frac{p_j}{m}\tG(z_2) \no \\
  & &-\frac{1}{8m^3c^2}\left(\sigma_i\sigma_j\tG(z_2)\bsig\cdot\op(z_2-V-\mu_B(\bsig\cdot{\bf B}_{\eff}))\bsig\cdot\op\tG(z_2)
  +\sigma_j\sigma_i\tG(z_1)\bsig\cdot\op(z_1-V-\mu_B(\bsig\cdot{\bf B}_{\eff}))\bsig\cdot\op\tG(z_1)\right) \no \\
  & &-\frac{1}{2m^3c^2}p_ip_j\left(\tG(z_1)+\tG(z_2)\right)
  -\frac{1}{4m^2c^2}\left(\sigma_i(z_1-V-\mu_B(\bsig\cdot{\bf B}_{\eff}))\sigma_j\tG(z_2)
  +\sigma_j(z_2-V-\mu_B(\bsig\cdot{\bf B}_{\eff}))\sigma_i\tG(z_1)\right) \no \\
  & &-\frac{i}{4m^3c^2}(z_1-z_2)\left(({\bf p }\times\bsig)_i\tG(z_1)p_j\tG(z_2)
  -p_i\tG(z_1)(\op\times\bsig)_j\tG(z_2)\right)\bigg\rangle_c,
\end{eqnarray}
where $H_{rc}$ and ${\bf v}_{rc}$ are the relativistic
corrections at order $1/c^2$ to the Hamiltonian and the velocity
respectively given by (\ref{RCH}) and (\ref{RCV}). The last term
on the right side hand does not contribute because
$(z_1-z_2)\rightarrow0$. Inserting the expression (\ref{lambda5})
in the conductivity (\ref{streda4}), we get $\tsig_{ij}^{I}$ at
order $1/c^2$
\begin{eqnarray}\label{sigmaI3}
  & &\tsig_{ij}^{I(2)}=\frac{e^2\hbar}{4\pi\Omega}
  {\mathrm Tr}\bigg\langle({\bf v}_{rc})_i(\tG^+-\tG^-)\frac{p_j}{m}\tG^-
  -({\bf v}_{rc})_i\tG^+\frac{p_j}{m}(\tG^+-\tG^-)+\frac{p_i}{m}(\tG^+-\tG^-)({\bf v}_{rc})_j\tG^- \no \\
  & &-\frac{p_i}{m}\tG^+({\bf v}_{rc})_j(\tG^+-\tG^-)
  +\frac{p_i}{m}(\tG^+H_{rc}\tG^+-\tG^-H_{rc}\tG^-)\frac{p_j}{m}\tG^-+\frac{p_i}{m}(\tG^+-\tG^-)\frac{p_j}{m}\tG^-H_{rc}\tG^- \no \\
  & &-\frac{p_i}{m}\tG^+H_{rc}\tG^+\frac{p_j}{m}(\tG^+-\tG^-)
  -\frac{p_i}{m}\tG^+\frac{p_j}{m}(\tG^+H_{rc}\tG^+-\tG^-H_{rc}\tG^-)\bigg\rangle_c \no \\
  & &-\varepsilon_{ijk}\frac{e^2\hbar}{16i\pi m^3c^2\Omega}{\mathrm Tr}\bigg\langle
  \left(\tG^+\bsig\cdot\op(\epsi_F-V-\mu_B(\bsig\cdot{\bf B}_{\eff}))\bsig\cdot\op\tG^+
  -\tG^-\bsig\cdot\op(\epsi_F-V-\mu_B(\bsig\cdot{\bf B}_{\eff}))\bsig\cdot\op\tG^-\right)\sigma_k\bigg\rangle_c \no \\
  & &+\frac{e^2\hbar}{16\pi
  m^2c^2\Omega}{\mathrm Tr}\bigg\langle(\tG^+-\tG^-)(\sigma_i(\epsi_F-V-\mu_B(\bsig\cdot{\bf B}_{\eff}))\sigma_j
  -\sigma_j(\epsi_F-V-\mu_B(\bsig\cdot{\bf B}_{\eff}))\sigma_i)\bigg\rangle_c.
\end{eqnarray}
The first term corresponds exactly to the relativistic corrections
that we get at oder $1/c^2$ in the Pauli approch. The two last
terms are supplementary terms which should not appear. We show
that they are cancelled by terms in $\tsig_{ij}^{II(2)}$. Indeed,
when we expand $D$ up to the second order in $1/c$ in the
expression (\ref{sigmaII2}) of $\tsig_{ij}^{II}$, we obtain
\begin{eqnarray}\label{sigmaII5}
  &&{\tilde\sigma}_{ij}^{II(2)}=-\frac{e^2}{4i\pi\Omega}{\mathrm Tr}\bigg\langle\left(\tG^+H_{rc}\tG^+-\tG^-H_{rc}\tG^-\right)
  \left(r_i\frac{p_j}{m}-r_j\frac{p_i}{m}\right)
  +\left(\tG^+-\tG^-\right)\left(r_i({\bf v}_{rc})_j-r_j({\bf v}_{rc})_i\right)\bigg\rangle_c \no \\
  & &+\varepsilon_{ijk}\frac{e^2\hbar}{16i\pi m^3c^2\Omega}
  {\mathrm Tr}\bigg\langle\left(\tG^+\bsig\cdot\op(\epsi_F-V-\mu_B(\bsig\cdot{\bf B}_{\eff}))\bsig\cdot\op\tG^+
  -\tG^-\bsig\cdot\op(\epsi_F-V-\mu_B(\bsig\cdot{\bf B}_{\eff}))\bsig\cdot\op\tG^-\right)\sigma_k\bigg\rangle_c \no \\
  & &-\frac{e^2\hbar}{16\pi
  m^2c^2\Omega}{\mathrm Tr}\bigg\langle(\tG^+-\tG^-)(\sigma_i(\epsi_F-V-\mu_B(\bsig\cdot{\bf B}_{\eff}))\sigma_j
  -\sigma_j(\epsi_F-V-\mu_B(\bsig\cdot{\bf B}_{\eff}))\sigma_i)\bigg\rangle_c.
\end{eqnarray}
The two last terms on the right hand side cancel the
supplementary terms in (\ref{sigmaI3}) and we obtain finally for
the total conductivity at order $1/c^2$
\begin{eqnarray}\label{totcond2}
  & &\tsig_{ij}^{(2)}=\tsig_{ij}^{I(2)}+\tsig_{ij}^{II(2)}=\frac{e^2\hbar}{4\pi\Omega}
  {\mathrm Tr}\left\langle({\bf v}_{rc})_i(\tG^+-\tG^-)\frac{p_j}{m}\tG^-\right.
  -({\bf v}_{rc})_i\tG^+\frac{p_j}{m}(\tG^+-\tG^-) \no \\
  & &+\frac{p_i}{m}(\tG^+-\tG^-)({\bf v}_{rc})_j\tG^--\frac{p_i}{m}\tG^+({\bf v}_{rc})_j(\tG^+-\tG^-)
  +\frac{p_i}{m}(\tG^+H_{rc}\tG^+-\tG^-H_{rc}\tG^-)\frac{p_j}{m}\tG^- \no \\
  & &+\left.\frac{p_i}{m}(\tG^+-\tG^-)\frac{p_j}{m}\tG^-H_{rc}\tG^-
  -\frac{p_i}{m}\tG^+H_{rc}\tG^+\frac{p_j}{m}(\tG^+-\tG^-)
  -\frac{p_i}{m}\tG^+\frac{p_j}{m}(\tG^+H_{rc}\tG^+-\tG^-H_{rc}\tG^-)\right\rangle_c \no \\
  & &-\frac{e^2}{4i\pi\Omega}{\mathrm Tr}\left\langle\left(\tG^+H_{rc}\tG^+-\tG^-H_{rc}\tG^-\right)
  \left(r_i\frac{p_j}{m}-r_j\frac{p_i}{m}\right)
  +\left(\tG^+-\tG^-\right)\left(r_i({\bf v}_{rc})_j-r_j({\bf
  v}_{rc})_i\right)\right\rangle_c.
\end{eqnarray}
}
This expression corresponds exactly to the one that we get from
the Pauli approach when we report the Pauli velocity at order
$1/c^2$, ${\bf v}_{rc}$, and the Pauli Hamiltonian at order
$1/c^2$, $H_{rc}$, in (\ref{streda4}) and (\ref{streda5}). For
higher order terms (say of order $1/c^{2n}$ with $n>1$), we can
predict that similar cancellations occur when we consider the
total conductivity $\tsig_{ij}^{I(2n)}+\tsig_{ij}^{II(2n)}$. We
have thus proved in this appendix the consistence between the
Pauli conductivity and the weak-relativistic limit of the Dirac
conductivity. \\

* Electronic address: crepieux@mpi-halle.mpg.de

\end{multicols}

\end{document}